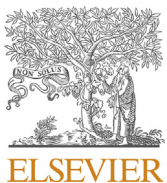
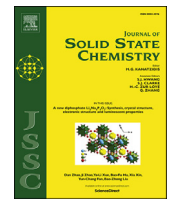

# Phase driven magnetic and optoelectronic properties of La$_2$CrNiO$_6$: A DFT and Monte Carlo perspective

Tushar Kanti Bhowmik [*], Tripurari Prasad Sinha

*Bose Institute, Department of Physics, 93/1, APC Road, Kolkata, 700009, India*

ABSTRACT

In search of the ferromagnetic insulators for spintronic device applications, we have studied the electronic, optical and magnetic properties of La$_2$CrNiO$_6$ (LCNO) using the first principle density functional theory (DFT) and the Monte Carlo simulation technique. Firstly, we have adopted the sol-gel method for the preparation of LCNO and refined the X-ray diffracted value using the Fullprof suite under the Rietveld mechanism. The obtained orthorhombic Pbnm and cubic Fm-3m space groups are used as the input of the DFT calculations. Both the structures have shown the half metallicity and insulting bandgap towards the majority and minority spin direction, respectively. From the magnetic moment calculations, we have seen the low spin state of Ni$^{3+}$, and the double exchange (DE) interaction are responsible for the ferromagnetic structure of LCNO. The optical dielectric constant is ten, and the refractive index is 3.2 for orthorhombic structure. The transition temperature (T$_C$) is determined to be 110 K from the MCS study, which has well agreed with experimental value (111 K). The RCP value is determined to 12 J/kg with a 5 T applied field, which is good for magnetocaloric applications.

## 1. Introduction

Double perovskites (A$_2$B′B″O$_6$, where, A is larger cation, B′ and B″ are smaller cations than A) have taken tremendous interest due to their versatile properties and applications. When the B-site is filled with 3d transition metals, it introduces various properties, such as different magnetic states [1–3], electron hopping conduction [4], ferromagnetic insulator [5–8], magnetodielectric [9], magnetocapacitance [10], magnetoresistance [11], the large magnetocaloric effect [12], etc. Most of these properties arise due to the BO$_6$ octahedra, where the B-site transition metal and oxygen transfers electron through the double exchange [13], superexchange mechanism [14]. Based on these properties, the double perovskites have been applied in storage devices [15], magnetic cooling technologies [12], spintronics [16,17], opto-electronics [18], piezoelectrics [19], etc. For spintronics devices, a ferromagnetic insulator and a non-magnetic conductor restrict electron transport through the non-magnetic region. Some of the ferromagnetic insulators which belongs to the double perovskites are La$_2$NiMnO$_6$ [5], La$_2$CoMnO$_6$ [6], La$_2$CrNiO$_6$ [20], Sr$_2$FeReO$_6$ [21] etc. The underlying physics behind these FM insulators is exciting due to their 3d and 5d orbitals. Again, the ferromagnetism always tries to accompany the metallic behaviour in the materials. So, alongside these materials, these two opposite natures proposed to study the underlying physics in detail.

First principle density functional theory (DFT) [22] is the most dependable technique to study these strongly correlated electron systems' energy bands and electronic structures. In DFT, the generalised gradient approximation (GGA) [23] along with the Coulomb potential U (GGA+U) [24] method is a widely accepted model to analyse the d and f electrons in strongly correlated systems. But this model is only studied to know the ground state properties of these materials. The Monte Carlo simulation technique has been used to study the thermally activated magnetic behaviour [25,26]. The detailed magnetic properties of some double perovskites have been studied previously through the Monte Carlo simulation [27–30]. There are several theoretical models, such as the 3D Ising model [31], Mean-field theory [32], 3D Heisenberg model [33], etc., to explain this behaviour appropriately.

At this point, we have chosen one of the double perovskite La$_2$CrNiO$_6$ (LCNO), which may be suitable for spintronics applications. LCNO is studied theoretically based on a computer-generated lattice parameters, which is unsuitable for this material [20]. So, we have revisited the complete theoretical studies with the actual lattice parameters and space group, which are determined experimentally. The electronic structure and detailed optical properties have been studied with two different symmetry cubic and orthorhombic, calculated from the Rietveld refinement [34]. After the detailed calculations, it is seen that the orthorhombic (Pbnm) symmetry is more close to experimental values than the cubic (Fm-3m) structure. The temperature-dependent magnetic studies from the Monte-Carlo simulation are well matched with the experimental one [35]. The input parameters for the Monte Carlo simulation are determined from the DFT calculations [36,37]. We have then calculated the magnetocaloric effect of LCNO using the Monte Carlo simulation [38]. The calculated relative cooling power (RCP) is very attractive for magnetic cooling as magnetocaloric material.

## 2. Experimental details

The Sol-Gel technique is applied to synthesize the double perovskite $La_2CrNiO_6$. The metallic nitrates $La(NO_3)_3.6H_2O$, $Cr(NO_3)_3.9H_2O$, and $Ni(NO_3)_2.6H_2O$ were taken with appropriate stoichiometric ratio and mixed in magnetic stirrer for 1 h with the de-ionised water medium. Then the citric acid and ethylene glycol are taken with proper molar ratio to the solution used as a fuelling and stabilizing agent, respectively. The whole solution is stirred for 5 h and then heated around 473 K to form the gel, and further heating caused an auto-combustion to achieve fluffy precursor LCNO powder. The obtained powder is then ground and calcined at 1173 K, and then it is pelletized with binding element polyvinyl chloride. Atlast, the sample is sintered at 1223 K to achieve the final LCNO powder [52]. The room temperature X-ray diffraction (XRD) pattern of LCNO is obtained in an X-ray powder diffractometer (Rigaku Miniflex II, Cu-K$\alpha$ : $\lambda$ = 1.54 Å), where the range of $2\theta$ is $10°$–$80°$ at a scanning rate of $0.02°$ per step. The crystal structure and the lattice parameters are extracted from the Rietveld refinement of the XRD data using the Fullprof suite program [39].

## 3. Computational details

### 3.1. Ab-initio calculations

The electronic structure and the magnetic properties of LCNO are being studied using the first principle density functional theory. The full-potential linearized augmented plane wave (FPLAPW) method has been implemented to solve the Kohn-Sham equations described in the WIEN2k package [40,41]. The crystal structure optimization and the electronic properties, and the magnetic moments of each ions have been calculated using the spin-polarized general gradient approximation (GGA). But, the GGA method has failed to separate the energy band near the Fermi level due to the strongly localized d-orbitals of Cr and Ni atoms [42]. So, we have introduced the GGA with the Coulomb repulsion U (GGA+U) to overcome this problem [43,44]. As the LCNO has shown two symmetric structures (discuss later) cubic and orthorhombic. The electronic and magnetic properties have been studied according to these symmetries. The muffin-tin raddi ($R_{mt}$) for La, Cr, Ni, and O atoms are 2.26 Å, 1.89 Å, 1.78 Å, and 1.69 Å for cubic, respectively. For orthorhombic, these radii are 2.26 Å, 1.89 Å, 1.78 Å, and 1.69 Å, respectively. The cut-off energy ($R_{mt} \times K_{max}$) from core states to valance states has been set to 6 and 7 eV, respectively. $17 \times 17 \times 17$ number of k-points have been adopted in the whole Brillion zone for self-consistent convergence for cubic phase and $10 \times 10 \times 8$ for the orthorhombic one. The Hubbard parameter $U_{eff} = U - J = 4$ eV and 6 eV have been used for Cr-d and Ni-d orbitals, respectively. The energy and charge convergence criteria have been fixed to $10^{-4}$ Ry and $10^{-3}$ e, respectively for self-consistent optimization calculations of LCNO. Structural optimization has been done to find the optimal lattice parameter to stabilize the crystal structures for both cases. After that, the different magnetic spin configuration are adopted, such as ferromagnetic (FM) and three type of antiferromagnetic (AFM) (A-AFM, C-AFM, G-AFM). Among them, the ferromagnetic configuration has the minimum energy. The individual atoms' magnetic moment is then calculated and discussed in detail in the following section. The detailed optical properties of LCNO for both structures have been analyzed from the optimize crystal structure.

The frequency dependent dielectric function $\epsilon(\omega)$ has been investigated to study the optical properties. It can be expressed as $\epsilon(\omega) = \epsilon_1(\omega) + j\epsilon_2(\omega)$, where $\epsilon_1(\omega)$ and $\epsilon_2(\omega)$ signify the real and imaginary component of the complex dielectric function $\epsilon(\omega)$. $\epsilon_2(\omega)$ is determined from the following equation.

$$\epsilon_2(\omega) = \frac{2e^2\pi}{\Omega\epsilon_0} \sum_{k,v,c}(|\psi_k^c|\mathbf{u}.\mathbf{r}|\psi_k^v|)^2 \delta(E_k^c - E_k^v - E) \quad (1)$$

Where $(|\psi_k^c|\mathbf{u}.\mathbf{r}|\psi_k^v|)^2$ represents the matrix elements for the direct transfer from the valance state to conduction state, $E_k^c - E_k^v$ is the excitation energy and $\psi_k^v$ and $\psi_k^c$ are the Bloch wave-functions for the valance and conduction band with the wave-vector k. The real component of the dielectric function is derived from the imaginary part $\epsilon_2(\omega)$ by using Kramers-Kronig relation,

$$\epsilon_2(\omega) = 1 + \frac{2}{\pi}P\int_0^\infty \frac{\omega'\epsilon_2(\omega')}{\omega'^2 - \omega^2}d\omega' \quad (2)$$

where $P$ is the principle value of the integral. Using the value of the dielectric functions, we have calculated other optical properties such as reflectivity ($R(\omega)$), refractive index ($n(\omega)$), absorption coefficient ($\alpha(\omega)$), and optical conductivity ($\sigma(\omega)$). These parameters are determined from the following equations, respectively.

$$R(\omega) = \left|\frac{\sqrt{\epsilon(\omega)} - 1}{\sqrt{\epsilon(\omega)} + 1}\right|^2 \quad (3)$$

$$n(\omega) = \frac{1}{\sqrt{2}}(\sqrt{\epsilon_1(\omega)^2 + \epsilon_2(\omega)^2} + \epsilon_1(\omega))^{\frac{1}{2}} \quad (4)$$

$$\alpha(\omega) = \sqrt{2}\omega(\sqrt{\epsilon_1(\omega)^2 + \epsilon_2(\omega)^2} - \epsilon_1(\omega))^{\frac{1}{2}} \quad (5)$$

$$\sigma(\omega) = -\frac{j\omega}{4\pi}\epsilon(\omega) \quad (6)$$

All of these parameters are calculated with a mesh of 1000 k-points over the irreducible Brillouin zone.

### 3.2. Monte Carlo simulations

The strongly correlated system has the strong anisotropy in spin configuration. For that reason, we have choosen the anisotropic Heisenberg spin model to do the Monte Carlo simulations. But For LCNO, the model has failed to satisfy the experimental data at low temperatures. So that we have tried with the discrete Ising spin model for the simulations, and it satisfactorily agreed with the experimental study. Actually, for LCNO there are two types of magnetic ions; the spin values (S) of both ions, $Cr^{3+}$ and $Ni^{3+}$, are 3/2. So that the anisotropy in the spin configuration vanishes itself due to double perovskite structure, which bounds themselves to keep the same number of unpair electrons in their 3d orbitals. However, we have taken the Ising model Hamiltonian for the simulation.

$$H = -J\sum_{<i,j>}s_is_j - h\sum_i s_i \quad (7)$$

Where $J$ and h are the interaction constant and the applied field, respectively. In our magnetic Monte Carlo simulation study, we have considered the magnetic $Cr^{3+}$ ion and $Ni^{3+}$ ion are situated next to each other, and the interaction between two magnetic ions is ferromagnetic. The coupling interaction $J$ is determined from the difference of the two magnetic configurations, FM and G-AFM [37]. The energy of these magnetic configurations is obtained earlier from the GGA+U method. We have calculated that the value of the interaction constant is 0.6 meV. The detailed calculation is given in the supplementary part.

We have performed the Monte Carlo simulation (MCS) using the single flip Metropolis algorithm to calculate the magnetic properties of LCNO. The periodic boundary condition and standard sampling method have been implemented to the whole lattice of size L = 30 in all three Cartesian directions. If the change in energy (dE) is less than zero, or the transition probability $[exp(-dE/k_BT)]$ is greater than a random value, the flip is accepted otherwise rejected. $10^6$ MCS steps have been used to reach the equilibrium of lattices and next $10^7$ steps to average the magnetization and other observables. The physical quantities, which have been measured using MCS, are described as follows [45]. The internal energy per site is,

$$E = \frac{1}{N} <H> \quad (8)$$

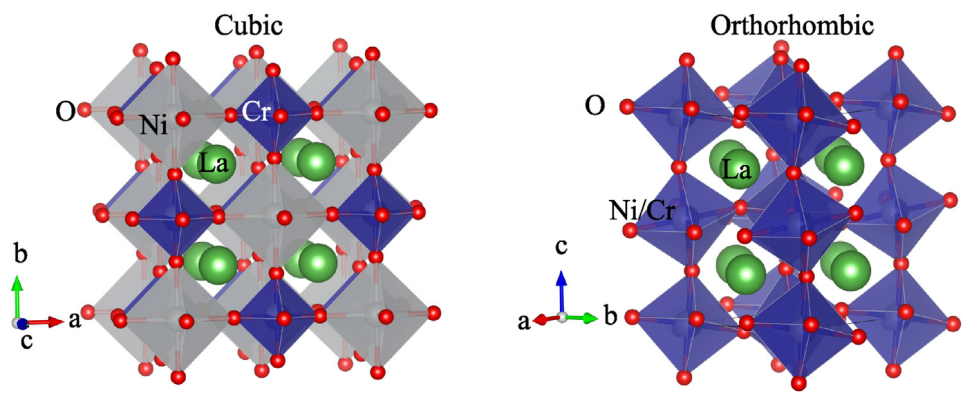

**Fig. 1.** The ordered cubic (left figure) and the disordered orthorhombic (right figure) octahedral structure of LCNO.

Where $N = L \times L \times L$ is the total number of lattice points. The magnetization of LCNO is,

$$M = <\frac{1}{N}\sum_i S_i> \quad (9)$$

The magnetic susceptibility is given by,

$$\chi = \frac{<M^2> - <M>^2}{k_BT} \quad (10)$$

The magnetic specific heat is calculated by,

$$C_m = \frac{<E^2> - <E>^2}{k_BT^2} \quad (11)$$

The magnetic entropy is given by

$$S_m = \int_0^T \frac{C_m}{T}dT \quad (12)$$

The change in magnetic entropy is described as

$$\Delta S_m = \int_0^{h_{max}} \left(\frac{\partial M}{\partial T}\right)dh \quad (13)$$

Where $h_{max}$ is the maximum applied field and $\left(\frac{\partial M}{\partial T}\right)_h$ is the thermal

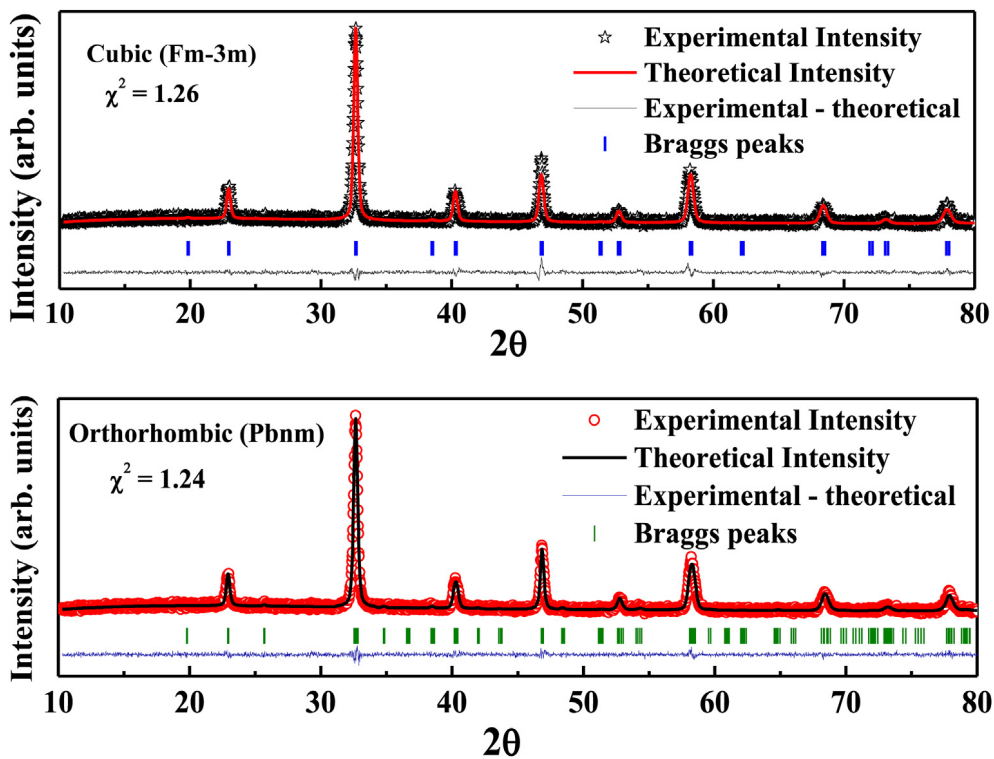

**Fig. 2.** The rietveld refinement of the experimental XRD data of cubic (upper) and orthorhombic (lower) LCNO. The description of colours have shown in figures. (For interpretation of the references to colour in this figure legend, the reader is referred to the Web version of this article.)

**Table 1**
Atomic coordinates and lattice parameters of LCNO. The Muffin-Tin radii ($R_{mt}$) of all atom used in WIEN2k described the right column of this table.

| Symmetry | Atom | x | y | z | Wyckoff positions | $R_{mt}$ (Å) |
|---|---|---|---|---|---|---|
| Cubic | | | | | | |
| Fm-3m | La | 0.2500 | 0.2500 | 0.2500 | 8c | 2.50 |
| ($\chi^2 = 1.26$) | Ni | 0.0000 | 0.0000 | 0.0000 | 4a | 2.02 |
| a = 7.7562 | Cr | 0.5000 | 0.0000 | 0.0000 | 4b | 1.85 |
| | O | 0.2844 | 0.0000 | 0.0000 | 24e | 1.67 |
| Orthorhombic | | | | | | |
| Pbnm | La | 0.4980 | 0.0164 | 0.2500 | 4c | 2.47 |
| ($\chi^2 = 1.24$) | Ni | 0.0000 | 0.0000 | 0.0000 | 4a | 1.86 |
| a = 5.5008 | Cr | 0.0000 | 0.0000 | 0.0000 | 4a | 1.82 |
| b = 5.4654 | O1 | 0.5555 | 0.4987 | 0.2500 | 4c | 1.60 |
| c = 7.7583 | O2 | 0.2131 | 0.2530 | 0.0441 | 8d | 1.60 |

magnetization for a fixed external magnetic field $h$. The adiabatic temperature change is given by

$$\Delta T_{ad} = -T \frac{\Delta S_m}{C_m} \quad (14)$$

The magnetocaloric effect is measured by a parameter called Relative Cooling Power (RCP), which is calculated from the magnetic entropy change vs temperature curves. The formula for determining of RCP is given by,

$$RCP = \int_{T_1}^{T_2} \Delta S_m(T) dT \quad (15)$$

Where, $T_1$ and $T_2$ are the cold and hot temperatures corresponding to both ends of half maximum value of $\Delta S_m$ vs $T$ curves.

## 4. Results and discussions

### 4.1. Crystallographic information

The Rietveld refinement of the XRD data is being performed to determine the crystal structure of LCNO. In this context, we have tried to fit the experimental data with orthorhombic Pbnm space group, as stated earlier [35]. The pseudo-voigt function is being imposed to match the peaks, whereas the base is fitted with the fifth order polynomial function, described in Fullprof software. A parameter that defines the good correlation with the experimental and theoretical results is called the goodness of fit ($\chi^2$). The minimum value of $\chi^2$ means a good refinement. In our case, it is 1.24, which turns out the difference curve with experimental and theoretical intensities to a straight line in Fig. 2. The fitted lower intensity peaks are shown in Supplementary Fig. S1. The higher angle peaks are well fitted for both the structures.

The stability of LCNO double perovskite maybe now approximated using the La–O, Ni–O, Cr–O bond lengths from the above refined data. It can be analyzed by Goldschmidt's tolerance factor [46], stated as

$$t = \frac{R_{La} + R_O}{\sqrt{2}\left(\frac{R_{Cr}+R_{Ni}}{2} + R_O\right)} \quad (16)$$

Where the $R_{La}$, $R_{Ni}$, $R_{Cr}$, $R_O$ are the ionic radii of La, Ni, Cr, O atoms, respectively. The ionic radii of these atoms are 1.16 Å, 0.69 Å, 0.61 Å, and 1.4 Å, respectively and the value of tolerance factor turns out to be 0.89, which is the limiting value of the orthorhombic phase. According to Goldschmidt, the value of tolerance factor should be 0.75 to 0.9 for orthorhombic and 0.9 to 1 for cubic phase. So, this tolerance value motivates us to further refinement of LCNO with cubic Fm-3m symmetry. Fig. 1(a) shows the XRD fitting with cubic phases, which indicates an excellent correlation with the experimental intensities. According to the Pbnm space group, the B-site cations are disordered and form randomly oriented B and B' sub-lattices. In the orthorhombic crystal structure, the Cr and Ni belong to the same 4a Wyckoff site. Whereas for cubic structure, these atoms have 4b and 4a sites, respectively. The lattice parameters and the atomic positions of both the symmetry group have been described in Table 1. The DFT calculations in the following sections discuss which one is preferable by nature between these symmetries.

### 4.2. Spin-polarized electronic structure

The spin-polarized electronic band structure plays a significant role in explaining the most of the physical properties of materials. The optical properties and charge transport mechanism largely depend on the energy bandgap of the semiconductors. The refined LCNO structure is optimized through the energy minimization process in WIEN2k. Here, we have taken four spin configurations, i.e., FM, G-AFM, A-AFM, and C-AFM. The result of the energy minimization suggests that the FM structure has the lowest energy than the others. The minimization curves are given in Supplementary Fig. S2. After optimizing the experimental lattice parameters and positions, the ground state band structure is calculated using the GGA+U method along the high symmetry direction within the Brillouin zone. In the up-spin channel, the band structure has shown a tiny spike above the Fermi energy at H-point in Fig. 3(a). So, the valance band maxima (VBM) crosses the Fermi energy ($E_F$) limit, which means the cubic Fm-3m structure possesses a metallic character in the ground state. In the minority spin channel, the value of the energy band gap is 3.4 eV. The VBM and CBM (conduction band minima) are in the same Γ point, indicating the direct bandgap present in this material. Fig. 3(b) represents the band structure of LCNO for the orthorhombic Pbnm space group. Here one of the conduction energy band is very close to the Fermi energy in the majority spin channel. The bandgap between VBM and CBM is 0.08 eV, which is negligibly small like silver selenide ($Ag_2Se$) [47]. So, the orthorhombic LCNO belongs to a narrow bandgap semiconductor category in the up-spin channel. Again, the down-spin channel has a wide direct band gap (Γ–Γ) 1.9 eV. The detailed study of the ground state electronic band structure justifies that LCNO belongs to a half-metallic perovskite family.

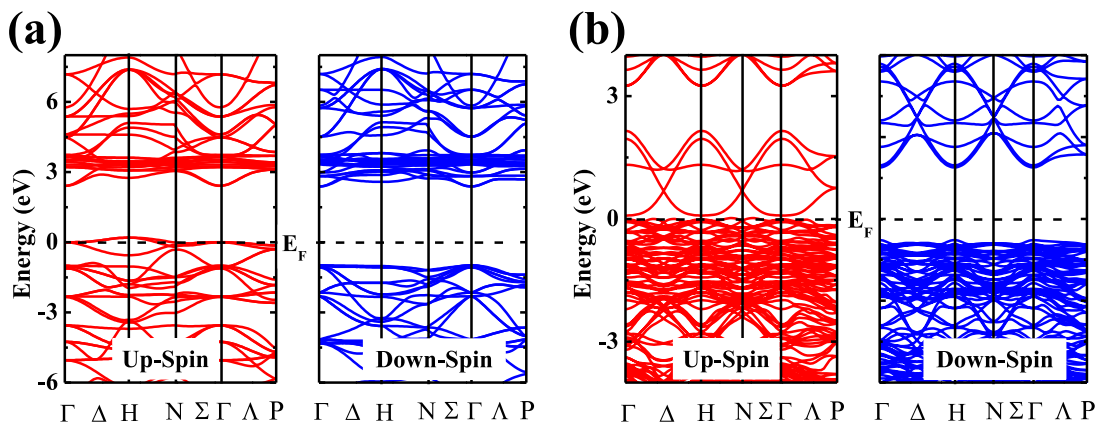

**Fig. 3.** The band structure plot of LCNO (a) cubic (b) orthorhombic. The Fermi energy ($E_F$) is set to zero, shown with the horizontal dotted line.

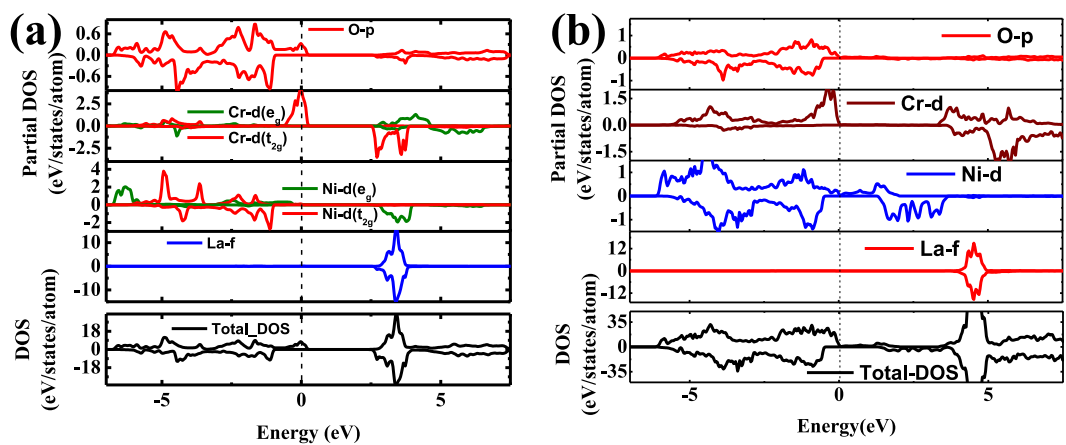

**Fig. 4.** The total and partial density of states of LCNO for (a) cubic and (b) orthorhombic structures.

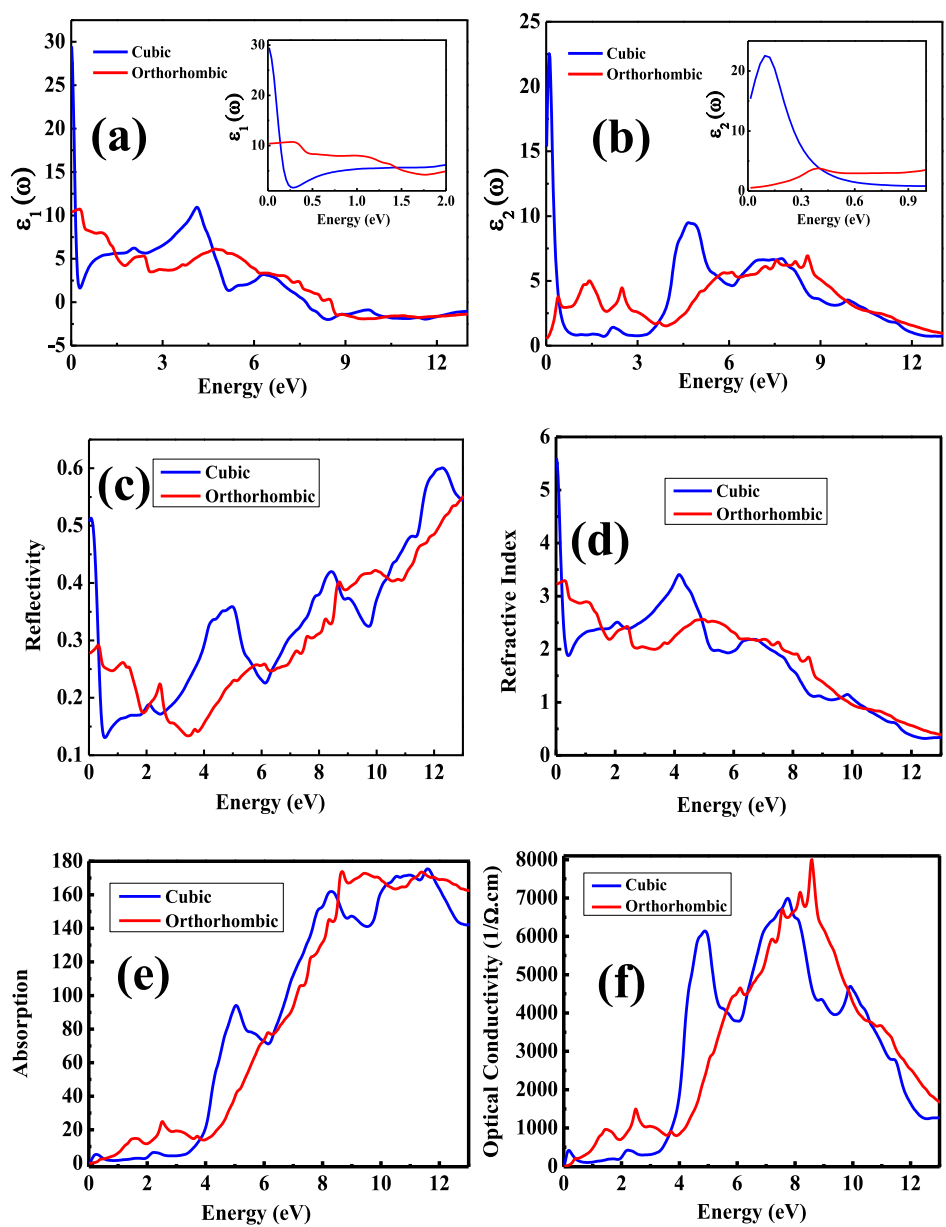

**Fig. 5.** (upper-left) The real part and (upper-right) imaginary part of the complex optical dielectric function, inset of both the figure are the larger view in the lower energy. The reflectivity (middle-left), refractive index (middle-right), absorption (lower-left) and optical conductivity (lower-right) of LCNO. Blue and red colour plots are described for the cubic and orthorhombic structure respectively. (For interpretation of the references to colour in this figure legend, the reader is referred to the Web version of this article.)

To understand the features of the band structures, the density of states (DOS) of all the atoms and the total DOS is studied in detail in this section. The DOS also predicts the nature of bonding as well as the electronic states of different energy levels. The partial and total DOS of all atoms have shown in Fig. 4 for both structures. Here the Fermi energy is set to zero electron volt. The left and right side of it corresponds valance electrons state and conduction electrons state, respectively. The positive and negative values of the states along the Y-axis signify the spin-up and spin-down channels, respectively. The valance band mainly consists of the Cr-3d and Ni-3d orbitals hybridized with the O-2p orbital for both structures. Similarly, the conduction band is dominated by the La-4f state and the 3d states of Cr and Ni atoms. The PDOS of La-f, which is symmetric with respect to the spin-up and down channel, signifies nonmagnetic. For cubic structure in the up-spin channel, the PDOS of Cr clearly shows that the Cr-d state (precisely Cr-d($t_{2g}$)) runs over the Fermi level. In the octahedral arrangement, the $t_{2g}$ orbital is occupied, which confirms the half-metallic behaviour of LCNO. The asymmetric DOS of Cr and Ni also support the ferromagnetic spin alignment of these atoms, which will be discussed in detail in the next section. The PDOS of Ni shows that there is no states near the Fermi level. In the orthorhombic structure, it is seen that the Cr-d state is below the Fermi energy, so the total DOS. But the Ni-d states is very close to the Fermi level. The maxima of the conduction states are shifted towards the higher energy from cubic to orthorhombic symmetry. The hybridization between La-4f and O-2p is very weak than the Cr-3d and O-2p, Ni-3d and O-2p. This corresponds the La–O bond is ionic type, whereas the Cr–O and Ni–O have a strong covalent bond.

*4.3. Optical properties*

The investigation of the optical properties is significant for material, as it may use in optoelectronic device applications. The intra band transitions are mainly responsible for generating various optical properties [53]. According to the Drude model, the transport properties depend on the charge-free carriers for the metals. The half-metallic ferromagnet shows the complete spin polarization, and it is optically transparent due to its metallic nature [48]. For that reason, various optical properties such as dispersion, absorption, polarization arise, which is useful for the device applications [49]. The energy-dependent optical parameters, which are mainly influenced by the energy band structures, inform us about the occupied and unoccupied states related to the energy bands. So, the real ($\epsilon_1(\omega)$) and imaginary ($\epsilon_2(\omega)$) part of the dielectric function have been determined from equations (1) and (2). Now, all the other optical and absorptive properties such as reflectivity, refractive index, absorption coefficient and optical conductivity have been calculated from the dielectric functions $\epsilon_1(\omega)$, $\epsilon_2(\omega)$ and $\epsilon(\omega)$. For cubic and orthorhombic symmetries, all the optical parameters with respect to the energy are described in Fig. 5(a–f) with blue and red colour, respectively. For cubic structure, we have seen only one non zero component, $\epsilon^{xx}(\omega) = \epsilon^{yy}(\omega) = \epsilon^{zz}(\omega)$ of the second order dielectric tensor. But in orthorhombic case, the tensor has three different non zero component along three axes. For simplification in the comparison between two structures, we have taken the average $[(\epsilon^{xx}(\omega) + \epsilon^{yy}(\omega) + \epsilon^{zz}(\omega))/3 = \epsilon^{avg}(\omega)]$ of these components and plotted in Fig. 5.

The real (dispersive) and imaginary (absorptive) components as the function of photon energy are plotted in Fig. 5(a) and (b), respectively.

The absorptive component $\epsilon_2(\omega)$ describes the optical transition mechanism between the energy bands in the materials. The transition between the highest occupied orbitals in the valance band and minimum unoccupied orbitals in conduction band produces the optical spectra in the materials. According to the selection rules the allowed optical transitions are $s \rightarrow p$, $p \rightarrow d$, $d \rightarrow p$, $d \rightarrow f$, etc. The first optical transitions are at 0.15 eV and 0.4 eV for cubic and orthorhombic structures, respectively [inset of Fig. 5(b)]. The values are significantly low, which justify the narrow bandgap of LCNO. Beyond the threshold, there are four significant transitions for orthorhombic at 1.5 eV, 2.5 eV, 6 eV and 9 eV. These transitions mainly from the O-2p orbital (VB) to Ni-3d, Cr-3d, La-5d (CB) and Ni-3d, Cr-3d to La-4f orbital. Similarly, there are three main transition at 2.25 eV, 5 eV and 7.5 eV for cubic structures, apart from the threshold one. The dispersive part $\epsilon_1(\omega)$ of dielectric function shows that it starts a maximum value at $\epsilon_1(0)$ and then creates some peaks and ends in the negative value for both the structures. The negative value of $\epsilon_1(\omega)$ indicates that the incident electromagnetic radiation is reflected from this medium in this energy range, which corresponds to the presence of the metallic character in LCNO. So, we can use this material for electromagnetic radiation protection purposes. The value of $\epsilon_1(\omega)$ at zero energy ($\omega = 0$) is known as the optical dielectric constant. For the cubic and orthorhombic symmetries, the values of $\epsilon_1(0)$ are 30 and 10, respectively.

Fig. 5(c) represents the reflectivity spectra in the photon energy range 0 eV–13 eV for both structures. The reflectivity is about to 50% at 0 eV for cubic symmetry and decreases sharply to a minimum value at 0.5 eV; after that, it varies from 15% to 50% up to 11 eV and reaches a maximum 60% at 12 eV. Where as the orthorhombic structure, the reflectivity goes to a minimum at 3.5 eV and reaches the maximum of 55% in this energy range. The refractive index $n(\omega)$ is an important parameter for determining the amount of light bent or refracted by the material. From Fig. 5(d), it is seen that the near $\omega = 0$ the $n(\omega)$ has a large value and decreases sharply in 0 eV–0.5 eV energy range due to the metallic nature in cubic symmetry. The $n(\omega)$ has reached a maximum value of 3.5 at 4 eV and decreased towards zero with increasing photon energy. But the orthorhombic symmetry, the static refractive index are determined 3.2, calculated at 0 eV. Above 10 eV, the value of $n(\omega)$ became less than the unity. This is called the superluminal phenomenon, where the speed of light in the material goes beyond the vacuum. This effect may be useful in optical networks, optical communications and optoelectronic applications.

The excitation energy of an electron for passing the bandgap of the material depends on the absorption of photons of light. The amount of power absorbed when the light is passing through a unit thickness of solid is called the absorption coefficient $\alpha(\omega)$. This $\alpha(\omega)$ is directly related to the imaginary component of dielectric function $\epsilon_2(\omega)$. In case of cubic symmetry (Fig. 5(e)), the absorption function has peaked at 2.2 eV, 5 eV and 8 eV, which correspond the inter-band transitions and the $\alpha(\omega)$ gradually takes the higher values with the high energy transitions. For orthorhombic structure, it has peaked at 1.5 eV, 2.5 eV and 6 eV, which are as same as the $\epsilon_2(\omega)$. The optical conductivity $\sigma(\omega)$ is described in Fig. 5(f), which shows three significant peaks for cubic structures. The orthorhombic symmetry of LCNO has two low energy peaks (1.6 eV and 2.2 eV), which correspond the electron transition in the visible range of the spectra. After reaching a maximum value, the $\sigma(\omega)$ decreases in the high energy (above 10 eV) for both cubic and orhorhomic structures of LCNO.

*4.4. Magnetic properties*

The magnetic properties of LCNO turn out to be a interesting one due to the interaction between the two magnetic atoms Cr and Ni, through the oxygen with the octahedral arrangement in the perovskite structure. The stable ground state magnetic structure is determined through the energy minimization of different magnetic structures to study the magnetic properties in detail. The ferromagnetic (FM) configuration has a lower energy value than the other three antiferromagnetic (AFM) spin orientations G-AFM, C-AFM and A-AFM. The magnetic moments of the individual atoms and the interstitial cubic and orthorhombic structures of LCNO are calculated using the DFT. The moments are described in Table 2 for both the structures. In the orthorhombic structures, the magnetic moment of Ni and Cr are 1.07 $\mu_B$ and 2.53 $\mu_B$, respectively. The spin only moment is 3.87 $\mu_B$ for both $Ni^{3+}$ as well as $Cr^{3+}$ ions. It is calculated from this formula stated as, $\mu = \sqrt{n(n+2)}$ (where n = number of unpair electrons in d-orbital). The experimental effective magnetic moment of LCNO is 4.01 $\mu_B$/f.u [35]. The total magnetic moment of LCNO is obtained 4.00 $\mu_B$/cell for orthorhombic structure, the same as the experimental reports by Palakkal et al. [35]. The obtained moment of $Ni^{3+}$ is significantly less than formulated value. In the octahedral arrangement of LCNO, the crystal field effect causes the 3d orbital splitting ($e_g$ and $t_{2g}$) of Ni and Cr atoms. Due to the stronger ligand field, the splitting energy ($\Delta$) of $Ni^{3+}$ becomes large enough to be a low spin state. So, the number of unpair electron in the 3d orbital turns out to be one. Goodenough et al. showed that in $LaNi_{0.5}Mn_{0.5}O_3$ perovskite, the spin only moment of low spin $Ni^{3+}$ has a value of 1 $\mu_B$, which is matched with our calculated result [50]. The reduction of the magnetic moment of $Cr^{3+}$ ion is due to the hybridization of 3d orbitals with the O-2p states. In the ordered cubic structure, the moment of Ni (1.6 $\mu_B$) is slightly greater whereas, Cr has lesser moment of 1.9 $\mu_B$. This happens because of the double exchange (DE) interaction, where one electron moves from O-2p orbital to the one unoccupied Ni-$e_g$ orbital and one Cr-$t_{2g}$ electron fills the gap in O-2p orbital. The DE interaction is mainly responsible for long-range FM ordering in LCNO.

**Table 2**
Calculated magnetic moments of each atoms, interstitial and cell for LCNO in $\mu_B$ unit.

| Symmetry | $\mu_{La}$ | $\mu_{Ni}$ | $\mu_{Cr}$ | $\mu_O$ | $\mu_{int}$ | $\mu_{cell}$ ($\mu_B$/cell) |
|---|---|---|---|---|---|---|
| Cubic | 0.01 | 1.69 | 1.96 | 0.009 | 0.26 | 3.99 |
| Orthorhombic | 0.01 | 1.07 | 2.53 | -0.001 | 1.50 | 4.00 |

Now, the temperature and field-dependent magnetization and the phase transition are studied thoroughly using the Monte Carlo simulation under the metropolis algorithm [51]. Fig. 6(a) represents the magnetization as a function of temperature at the different magnetic fields. Upon cooling, the value of the magnetization suddenly increases and reaches a maximum of 1.7 $\mu_B/at$. This type of behaviour is called the ferromagnetic phase transition. The magnetic susceptibility ($\chi$) is plotted in Fig. 6(b) to determine the exact transition temperature ($T_C$). The peak in the susceptibility curve at 110 K suggested the $T_C$ of LCNO. The steepness of the magnetization versus temperature curve gradually decreases with the applied field. The alignment of the spin increases with the applied field in the paramagnetic region causes the increment of magnetization.

### 4.5. Magnetocaloric effect

The magnetocaloric effect (MCE) is nothing but the heating as well as cooling of the magnetic materials under the application of an external magnetic field. The range of heating or cooling defines the effectiveness of the magnetic materials. The MCE is characterized by the entropy change ($\Delta S$) and the temperature change ($\Delta T$) during the isothermal and adiabatic processes, respectively. The adiabatic temperature change ($\Delta T_{ad}$) with respect to the temperature at the different magnetic fields is shown in the upper inset of Fig. 7. The maximum change is recorded at the transition temperature, and value of the maxima increases with a higher magnetic field. The FWHM value of this curve describes the amount of cooling in the materials. The value of the $\Delta T_{ad}^{max}$ is 33 K with 5 T applied field. The negative magnetic entropy change $\Delta S_M$ is shown in the lower inset of Fig. 7. It is negative in the entire temperature range. The peak value gradually increases from 0.05 J/kg.K to 0.24 J/kg.K with applied field change from 1 T to 5 T. The operating temperature range for cooling of the material determines from the FWHM of the $\Delta S_M$ versus T (K) curve. For this purpose, the RCP is calculated at the different applied fields, and it is proportional to the field. The RCP value is 12 J/kg at 5 T, which is good for magnetocaloric applications.

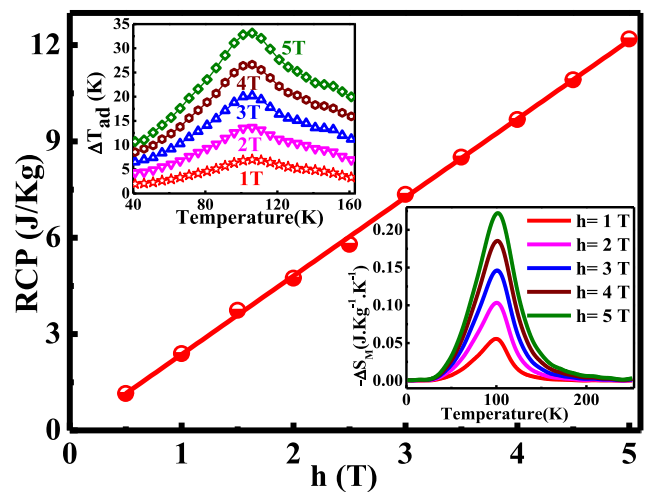

**Fig. 7.** The RCP at different field, which varies linearly upto 5 T magnetic field. (upper-inset) Adiabatic temperature change with respect to the temperature at different magnetic field, (lower-inset) The isothermal entropy change at different applied field.

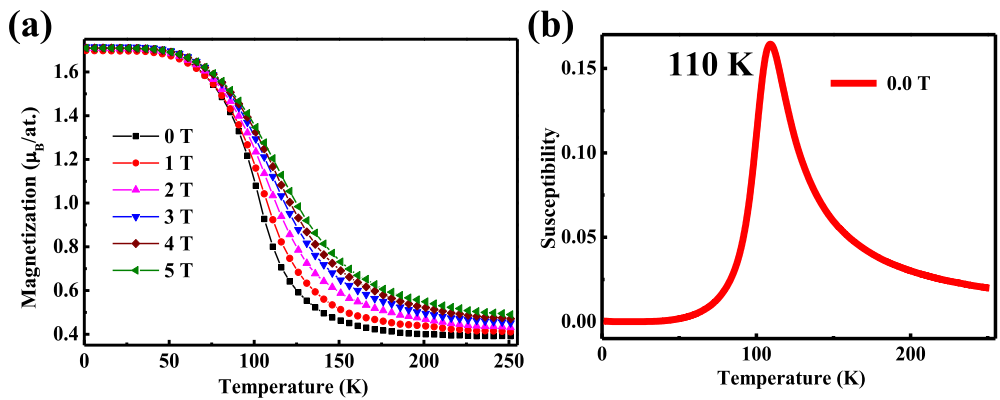

**Fig. 6.** (a) Magnetization vs temperature plot at different applied field. (b) Magnetic susceptibility plot at zero applied field.

## 5. Conclusions

We have studied the electronic, optical and magnetic properties of LCNO using the first principle DFT calculations and reinvestigated the magnetic properties and the magnetocaloric effect using the Monte Carlo simulation method. Some theoretical studies have reported earlier without the experimental evidence on the basis of unrealistic structures. But, we have done all the calculations from the experimental XRD structure, and the obtained results have followed the experimental one. From the XRD refinement, we have seen that there are two symmetries (orthorhombic- Pbnm and cubic- Fm-3m), which have fitted nicely with the experimental one. According to these structures, we have investigated these properties and compared with the experiment. From the band structures and DOS of both structures, we have concluded that LCNO has half-metallic properties and magnetic properties confirm that it is ferromagnetic also. It has high dielectric constant and refractive index. All these properties suggest that it is a very good candidate for spintronic applications. We have conclude that the orthorhombic Pbnm space group is more experimentally accepted than the cubic one. Needless to say, the magnetocaloric study also shows that it has a good RCP value for magnetic refrigeration.

## CRediT authorship contribution statement

**Tushar Kanti Bhowmik:** Conceptualization, Experiment, Methodology, Software, Formal analysis, Analysis, Investigation, Writing – review & editing. **Tripurari Prasad Sinha:** Supervision.

## Declaration of competing interest

The authors declare that they have no known competing financial interests or personal relationships that could have appeared to influence the work reported in this paper.

## Acknowledgement

The authors thank Tanay Dey and Suvajit Patra for their valuable discussion over Monte-Carlo code. T.K Bhowmik would like to thank Department of Science and Technology (DST), Government of India for providing the financial support in the form of DST-INSPIRE fellowship (IF160418).

## Appendix A. Supplementary data

Supplementary data to this article can be found online at https://doi.org/10.1016/j.jssc.2021.122570.

## References

[1] D. Serrate, J.M.D. Teresa, M.R. Ibarra, Double perovskites with ferromagnetism above room temperature, J. Phys. Condens. Matter 19 (2006), 023201.

[2] Y. Moritomo, T. Akimoto, A. Nakamura, K. Ohoyama, M. Ohashi, Antiferromagnetic metallic state in the heavily doped region of perovskite manganites, Phys. Rev. B 58 (1998) 5544–5549.

[3] W. Prellier, V. Smolyaninova, A. Biswas, C. Galley, R.L. Greene, K. Ramesha, J. Gopalakrishnan, Properties of the ferrimagnetic double perovskites $A_2FeReO_6$ (A = Ba and Ca), J. Phys. Condens. Matter 12 (2000) 965–973.

[4] S. Halder, M. Sheikh, B. Ghosh, T. Sinha, Electronic structure and electrical conduction by polaron hopping mechanism in $A_2LuTaO_6$ (A= Ba, Sr, Ca) double perovskite oxides, Ceram. Int. 43 (2017) 11097–11108.

[5] H. Das, U.V. Waghmare, T. Saha-Dasgupta, D.D. Sarma, Electronic structure, phonons, and dielectric anomaly in ferromagnetic insulating double perovskite $La_2NiMnO_6$, Phys. Rev. Lett. 100 (2008) 186402.

[6] S. Baidya, T. Saha-Dasgupta, Electronic structure and phonons in $La_2CoMnO_6$: a ferromagnetic insulator driven by coulomb-assisted spin-orbit coupling, Phys. Rev. B 84 (2011), 035131.

[7] B. Kim, J. Lee, B.H. Kim, H.C. Choi, K. Kim, J.-S. Kang, B.I. Min, Electronic structures and magnetic properties of a ferromagnetic insulator: $La_2MnNiO_6$, J. Appl. Phys. 105 (2009), 07E515.

[8] W. Yi, Y. Matsushita, A. Sato, K. Kosuda, M. Yoshitake, A.A. Belik, $Bi_3Cr_{2.91}O_{11}$: a ferromagnetic insulator from $Cr^{4+}/Cr^{5+}$ mixing, Inorg. Chem. 53 (2014) 8362–8366. PMID: 25089932.

[9] H. Lin, X.X. Shi, X.M. Chen, $LaRNiMnO_6$ (R = Pr, Nd, Sm) double perovskites with magnetodielectric effects, J. Alloys Compd. 709 (2017) 772–778.

[10] N. Rogado, J. Li, A. Sleight, M. Subramanian, Magnetocapacitance and magnetoresistance near room temperature in a ferromagnetic semiconductor: $La_2NiMnO_6$, Adv. Mater. 17 (2005) 2225–2227.

[11] J. Yang, J. Kim, Y. Woo, C. Kim, B. Lee, Magnetoresistance in double perovskites $Ba_{2-x}La_xFeMoO_6$, J. Magn. Magn Mater. 310 (2007) e664–e665. Proceedings of the 17th International Conference on Magnetism.

[12] J.Y. Moon, M.K. Kim, Y.J. Choi, N. Lee, Giant anisotropic magnetocaloric effect in double-perovskite $Gd_2CoMnO_6$ single crystals, Sci. Rep. 7 (2017) 16099.

[13] H.-Z. Lin, C.-Y. Hu, P.-H. Lee, A.Z.-Z. Yan, W.-F. Wu, Y.-F. Chen, Y.-K. Wang, Half-metallic property induced by double exchange interaction in the double perovskite $Bi(2)BB'O(6)$ (B, B' = 3d Transitional Metal) via first-principles calculations, Materials (Basel, Switzerland) 12 (2019) 1844, 31174337[pmid].

[14] H. Das, P. Sanyal, T. Saha-Dasgupta, D.D. Sarma, Origin of magnetism and trend in $T_c$ in Cr-based double perovskites: interplay of two driving mechanisms, Phys. Rev. B 83 (2011) 104418.

[15] H. Nair, P. Ramachandran, S. Venkataraman, S. K, Exchange bias and memory effect in double perovskite $Sr_2FeCoO_6$, Appl. Phys. Lett. 101 (2012) 142401.

[16] J. Xu, C. Liu, J.-B. Liu, L. Bellaiche, H. Xiang, B.-X. Liu, B. Huang, Prediction of room-temperature half-metallicity in layered halide double perovskites, npj Comput. Mater. 5 (2019) 114.

[17] S.-J. Chang, M.-H. Chung, M.-Y. Kao, S.-F. Lee, Y.-H. Yu, C.-C. Kaun, T. Nakamura, N. Sasabe, S.-J. Chu, Y.-C. Tseng, $GdFe_{0.8}Ni_{0.2}O_3$: a multiferroic material for low-power spintronic devices with high storage capacity, ACS Appl. Mater. Interfaces 11 (2019) 31562–31572.

[18] Y. Fu, H. Zhu, J. Chen, M.P. Hautzinger, X.-Y. Zhu, S. Jin, Metal halide perovskite nanostructures for optoelectronic applications and the study of physical properties, Nat. Rev. Mater. 4 (2019) 169–188.

[19] Y.-M. You, W.-Q. Liao, D. Zhao, H.-Y. Ye, Y. Zhang, Q. Zhou, X. Niu, J. Wang, P.-F. Li, D.-W. Fu, Z. Wang, S. Gao, K. Yang, J.-M. Liu, J. Li, Y. Yan, R.-G. Xiong, An organic-inorganic perovskite ferroelectric with large piezoelectric response, Science 357 (2017) 306–309.

[20] J. Wang, X. Hao, Y. Xu, Z. Li, N. Zu, Z. Wu, F. Gao, Cation ordering induced semiconductor to half metal transition in $La_2NiCrO_6$, RSC Adv. 5 (2015) 50913–50918.

[21] A.K. Paul, M. Reehuis, V. Ksenofontov, B. Yan, A. Hoser, D.M. Többens, P.M. Abdala, P. Adler, M. Jansen, C. Felser, Lattice instability and competing spin structures in the double perovskite insulator $Sr_2FeOsO_6$, Phys. Rev. Lett. 111 (2013) 167205.

[22] W. Kohn, L.J. Sham, Self-consistent equations including exchange and correlation effects, Phys. Rev. 140 (1965). A1133–A1138.

[23] J.P. Perdew, K. Burke, M. Ernzerhof, Generalized gradient approximation made simple, Phys. Rev. Lett. 77 (1996) 3865–3868.

[24] V.I. Anisimov, F. Aryasetiawan, A.I. Lichtenstein, First-principles calculations of the electronic structure and spectra of strongly correlated systems: the LDA + U method, J. Phys. Condens. Matter 9 (1997) 767–808.

[25] R. Masrour, L. Bahmad, E.K. Hlil, M. Hamedoun, A. Benyoussef, Superparamagnetic behavior in $la_{0.7}ca_{0.3}mno_3$ perovskite: Monte Carlo simulations, J. Supercond. Nov. Magnetism 28 (2015) 165–168.

[26] S. Idrissi, H. Labrim, S. Ziti, R. Khalladi, N. El Mekkaoui, I. El Housni, S. Mtougui, L. Bahmad, Magnetic properties of the double perovskite $bi_2fecro_6$, J. Electron. Mater. 48 (2019) 3579–3587.

[27] A. Nid-bahami, A. El Kenz, A. Benyoussef, L. Bahmad, M. Hamedoun, H. El Moussaoui, Magnetic properties of double perovskite $sr_2ruhoo_6$: Monte Carlo simulation, J. Magn. Magn Mater. 417 (2016) 258–266.

[28] S. Idrissi, R. Khalladi, S. Mtougui, S. Ziti, H. Labrim, I. El Housni, N. El Mekkaoui, L. Bahmad, Magnetism and phase diagrams of the doubles perovskite $sr_2criro_6$: Monte Carlo simulations, Phys. Stat. Mech. Appl. 523 (2019) 714–722.

[29] M. Arejdal, L. Bahmad, A. Abbassi, A. Benyoussef, Magnetic properties of the double perovskite $ba_2niuo_6$, Phys. Stat. Mech. Appl. 437 (2015) 375–381.

[30] M. El Yadari, L. Bahmad, A. El Kenz, A. Benyoussef, Monte Carlo study of the double perovskite nano $sr_2vmoo_6$, J. Alloys Compd. 579 (2013) 86–91.

[31] A.L. Talapov, H.W.J. Blote, The magnetization of the 3D Ising model, J. Phys. Math. Gen. 29 (1996) 5727–5733.

[32] G. Toulouse, M. Gabay, Mean field theory for heisenberg spin glasses, J. Phys. Lattr. 42 (1981) 103–106.

[33] C. Holm, W. Janke, Critical exponents of the classical three-dimensional heisenberg model: a single-cluster Monte Carlo study, Phys. Rev. B 48 (1993) 936–950.

[34] H.M. Rietveld, A profile refinement method for nuclear and magnetic structures, J. Appl. Crystallogr. 2 (1969) 65–71.

[35] J.P. Palakkal, T. Faske, M. Major, I. Radulov, P. Komissinskiy, L. Alff, Ferrimagnetism, exchange bias and spin-glass property of disordered $La_2CrNiO_6$, J. Magn. Magn Mater. 508 (2020) 166873.

[36] M. Arejdal, M. Kadiri, A. Abbassi, A. Slassi, A.A. Raiss, L. Bahmad, A. Benyoussef, Magnetic properties of the doubleperovskite $ba_2couo_6$: ab initio method, mean field approximation, and Monte Carlo study, J. Supercond. Nov. Magnetism 29 (2016) 2659–2667.

[37] S. Sidi Ahmed, M. Boujnah, L. Bahmad, A. Benyoussef, A. El Kenz, Magnetic and electronic properties of double perovskite $Lu_2MnCoO_6$: ab-initio calculations and Monte Carlo simulation, Chem. Phys. Lett. 685 (2017) 191–197.

[38] K. Binder, D. Heermann, L. Roelofs, A.J. Mallinckrodt, S. McKay, Monte Carlo simulation in statistical physics, Comput. Phys. 7 (1993) 156–157.

[39] J. Rodríguez-Carvajal, Recent advances in magnetic structure determination by neutron powder diffraction, Phys. B Condens. Matter 192 (1993) 55–69.

[40] P. Blaha, K. Schwarz, P. Sorantin, S. Trickey, Full-potential, linearized augmented plane wave programs for crystalline systems, Comput. Phys. Commun. 59 (1990) 399–415.

[41] P. Blaha, K. Schwarz, F. Tran, R. Laskowski, G.K.H. Madsen, L.D. Marks, Wien2k: an apw+lo program for calculating the properties of solids, J. Chem. Phys. 152 (2020), 074101.

[42] S.A. Khandy, J.-D. Chai, Thermoelectric properties, phonon, and mechanical stability of new half-metallic quaternary heusler alloys: Ferhcrz (z = si and ge), J. Appl. Phys. 127 (2020) 165102.

[43] S. Ahmad Khandy, J.-D. Chai, Robust stability, half-metallic ferrimagnetism and thermoelectric properties of new quaternary heusler material: a first principles approach, J. Magn. Magn Mater. 502 (2020) 166562.

[44] S.A. Khandy, D.C. Gupta, Electronic structure, magnetism and thermoelectric properties of double perovskite sr$_2$honbo$_6$, J. Magn. Magn Mater. 458 (2018) 176–182.

[45] K. Binder, D. Heermann, Monte Carlo Simulation in Statistical Physics : An Introduction 80, 2010.

[46] V.M. Goldschmidt, Die gesetze der krystallochemie, Naturwissenschaften 14 (1926) 477–485.

[47] F. Kirchhoff, J.M. Holender, M.J. Gillan, Structure, dynamics, and electronic structure of liquid ag-se alloys investigated by ab initio simulation, Phys. Rev. B 54 (1996) 190–202.

[48] Q. Mahmood, M. Hassan, M. Yaseen, A. Laref, Half-metallic ferromagnetism and optical behavior in alkaline-earth metals based beryllium perovskites: Dft calculations, Chem. Phys. Lett. 729 (2019) 11–16.

[49] T. Zhang, J.-H. Lin, Y.-M. Yu, X.-R. Chen, W.-M. Liu, Stacked bilayer phosphorene: strain-induced quantum spin hall state and optical measurement, Sci. Rep. 5 (2015) 13927.

[50] A. Wold, R.J. Arnott, J.B. Goodenough, Some magnetic and crystallographic properties of the system LaMn$_{(1-x)}$Ni$_x$O$_{3+\lambda}$, J. Appl. Phys. 29 (1958) 387–389.

[51] T.K. Bhowmik, T.P. Sinha, Al-dependent electronic and magnetic properties of YCrO$_3$ with magnetocaloric application: an ab-initio and Monte Carlo approach, Phys. B Condens. Matter 606 (2021) 412659.

[52] T.K. Bhowmik, M.S. Sheikh, A.P. Sakhya, A. Dutta, T.P. Sinha, Synthesis, structural and electrical conductivity of half-metallic perovskite oxide La$_2$CrNiO$_6$, AIP Conf. Proc. 2369 (2021), 020080.

[53] S. Halder, T.K. Bhowmik, A. Dutta, T.P. Sinha, The photophysical anisotropy and electronic structure of new narrow band gap perovskites Ln$_2$AlMnO$_6$ (Ln = La, Pr, Nd): An experimental and DFT perspective, Ceram. Int. 46 (2020) 21021.